\documentclass{article}
\usepackage{amsfonts}
\usepackage{epsfig,amsmath}
\usepackage{natbib}
\usepackage{amssymb}

\newcommand{\corr}[1]{\ensuremath{\mathrm{Corr}[#1]}}
\newcommand{\bd}[1]{\ensuremath{\mbox{\boldmath $#1$}}}

\begin{document}

\title{Using prior information to identify boundaries in disease risk maps}

\author{Duncan Lee\\
School of Mathematics and Statistics, University of Glasgow, Glasgow, UK}

\maketitle

\begin{abstract}
{Disease maps display the spatial pattern in disease risk, so that high-risk clusters can be identified. The spatial structure in the risk map is typically represented by a set of random effects, which are modelled with a conditional autoregressive (CAR) prior. Such priors include a global spatial smoothing parameter, whereas real risk surfaces are likely to include areas of smooth evolution as well as discontinuities, the latter of which are known as risk boundaries. Therefore, this paper proposes an extension to the class of CAR priors, which can identify both areas of localised spatial smoothness and risk boundaries. However, allowing for this localised smoothing requires large numbers of correlation parameters to be estimated, which are unlikely to be well identified from the data. To address this problem we propose eliciting an informative prior about the locations of such boundaries, which can be combined with the information from the data to provide more precise posterior inference. We test our approach by simulation, before applying it to a study of the risk of emergency admission to hospital in Greater Glasgow, Scotland.\\}

\end{abstract}

%%%%%%%%%%
%%%% Introduction
%%%%%%%%%%
\section{Introduction}
Disease maps display the spatial pattern in disease risk over a city or country, which enables areas at high risk to be identified. The identification of such areas informs government policy on health provision, such as which sub-populations should be targeted for an educational campaign about key risk factors, or where should a new hospital be built. Disease maps are produced using population level disease counts relating to small non-overlapping areal units, because the exact locations of individual cases are often not available for confidentiality reasons. Thus, these maps display population level summaries of disease risk, rather than relating to specific individuals.  The area of disease mapping is now well established, and is the subject of numerous books (for example \cite{elliott2000} and \cite{lawson2008}) and research articles (for example \cite{knorrheld2000}, \cite{macnab2003}, \cite{wakefield2007} and \cite{lee2011}). The growth in this research area is due in part to the wide availability of the required data, with examples including the Surveillance Epidemiology and End Results (SEER) cancer database in the USA, and the neighbourhood statistics databases in the UK.\\

Bayesian hierarchical models are typically used to produce disease maps, which represent the spatial pattern in disease risk by a vector of covariates and a set of random effects.  The latter act as a surrogate for unknown or unmeasured covariates, which induce spatial correlation into the disease data but cannot be modelled directly. Typically, Conditional Autoregressive (CAR, \cite{besag1974}) models are specified for the random effects,  with examples including the commonly used BYM model (\cite{besag1991}), as well as alternatives  developed by \cite{leroux1999} and \cite{stern1999}.  The spatial correlation structure induced by CAR models is determined by geographical adjacency, so that if two areas share a common border their random effects are correlated, otherwise they are conditionally independent given the values of the remaining random effects.\\

These models thus provide a global representation of spatial correlation, because the  random effects can range from being spatially smooth everywhere through to independent everywhere. However, this global interpretation is likely to be overly simplistic for real disease data, which are instead likely to contain areas of strong spatial correlation as well as locations at which abrupt step changes occur.  Such step changes may occur where different communities, such as those that are rich and poor, border each other, and are therefore likely to have very different disease risks despite sharing a common border. An example of this phenomena is provided by Figure \ref{figure SIR}, which displays the standardised incidence ratio (SIR) for emergency admissions to hospital in Greater Glasgow, Scotland during 2010. The map exhibits highly localised spatial structure, and suggests that the existing global CAR models are unlikely to adequately capture the spatial pattern in these data. Furthermore, the identification of such neighbourhood boundaries are of interest in their own right in a sociological context, because their locations may reflect changes in the biological, physical or social processes underlying the response (\cite{jacquez2000}).\\

A number of extensions have been proposed to CAR priors to accommodate localised spatial smoothing and the identification of boundaries, including recent papers by \cite{lu2007}, \cite{ma2010},  \cite{li2011}, \cite{lee2012} and \cite{lee2012b}.  The main idea underlying these extensions is to jointly model the  spatial adjacency structure of the random effects  in addition to the remaining model parameters, rather than basing the former on geographical adjacency. Specifically, pairs of random effects that relate to geographically adjacent areas can be modelled as either conditionally independent or correlated, rather than having correlation enforced upon them. However, the main statistical challenge with these extensions  is the large numbers of additional parameters to be estimated, which far outnumber the data and are thus likely to be only weakly identifiable. For example, in the Greater Glasgow region considered in Section 5 there are 271 data points and 701 parameters that control the spatial adjacency structure of the random effects.\\

Therefore, in this paper we propose augmenting the information from the data with an informative prior distribution, which should lead to more informative posterior inference about the localised spatial structure of the random effects. However, obtaining such prior information is far from straightforward, and we propose eliciting it from a second data set comprising disease counts for an earlier time period. Due to the large availability of small-area statistics data from multiple years are typically available, and should exhibit very similar spatial structure to the response (unless large-scale re-generation has taken place between the two time periods). To assess the utility of this approach we compare it to using a number of weakly informative priors, which provide little information about the spatial structure of the random effects.\\

The remainder of this paper is structured as follows. Section two provides a brief review of Bayesian disease mapping, while Section three outlines our methodological innovation, including the elicitation of an informative prior distribution. In Section 4 the efficacy of our approach is assessed via simulation, while in Section 5 it is applied to emergency hospital admissions data in Greater Glasgow in 2010.  Finally, the paper ends with a concluding discussion in Section 6, highlighting avenues for future work.\\

%%%%%%%%%%
%%%% Background
%%%%%%%%%%
\section{Bayesian disease mapping}

\subsection{Notation and likelihood model}
The region under study is partitioned into $n$ non-overlapping areal units $A_{1},\ldots, A_{n}$, which are typically administrative districts such as electoral wards or census tracts. The response variable is a  vector of disease counts denoted by $\mathbf{Y}=(Y_{1},\ldots,Y_{n})$, where $Y_k$ is the number of cases of the disease in question in areal unit $k$ in a specified time period (e.g. a year). In addition, a vector $\mathbf{E}=(E_{1},\ldots,E_{n})$ comprising the expected numbers of disease cases is also available, which is computed by external standardisation and used to adjust for the varying population demographics in each area. Using these data the simplest measure of disease risk is the standardised incidence ratio, which for area $k$ is computed as $\mbox{SIR}_k=Y_k/E_k$.  However, this is an unstable estimate of disease risk, especially if the expected numbers of cases are small. Therefore, disease risk is more commonly estimated using a Bayesian hierarchical model, the likelihood component of which is given by

\begin{eqnarray}
Y_{k}|E_k,R_k &\sim&\mbox{Poisson}(E_kR_k)~~~~\mbox{for }k=1,\ldots,n,\nonumber\\
\ln(R_{k})&=&\mathbf{x}_{k}^{\tiny\mbox{T}\normalsize}\bd{\beta} + \phi_{k}.\label{equation likelihood}
\end{eqnarray}

Disease risk in area $k$ is denoted by $R_k$, and is modelled by a vector of $p$ covariates
$\mathbf{x}_{k}=(1, x_{k1}, \ldots, x_{kp})$ (including an intercept term) and a random effect $\phi_k$, the latter of which allows for over dispersion and spatial correlation in the disease data caused by the existence of unmeasured or unknown confounders. The corresponding regression parameters $\bd{\beta}=(\beta_0,\beta_1,\ldots,\beta_p)$ are assigned independent Gaussian priors, where a large variance is typically chosen to make them weakly informative.

\subsection{Random effects model}
The random effects  $\bd{\phi}=(\phi_{1},\ldots,\phi_{n})$ are typically assigned a conditional autoregressive prior, which is a special case of a Gaussian Markov Random Field (GMRF). A general representation of this class of models is $\bd{\phi}\sim\mbox{N}(\mathbf{0},\tau^{2}Q(W)^{-1})$, where $Q(W)$ is a sparse precision matrix that may or may not be invertible. The sparsity comes from a binary $n\times n$ neighbourhood or adjacency matrix $W$, where $w_{kj}$ equals one if areal units $(k,j)$ share a common border (denoted $k\sim j$) and is zero otherwise (denoted $k\nsim j$). In the latter case $Q(W)_{kj}$ also equals zero, and $(\phi_{k},\phi_{j})$ are conditionally independent given the values of the remaining random effects. A number of CAR priors have been proposed in the disease mapping literature, and a comparison and review is provided by \cite{lee2011}. One such model was proposed by \cite{leroux1999}, which has precision matrix:

\begin{equation}
Q(W, \rho)~=~[\rho(\mbox{diag}(w_{k+}) - W) + (1-\rho)I].\label{equation lerouxjoint}
\end{equation}

In the above equation $\rho$ is a global spatial correlation or smoothing parameter, while $\mbox{diag}(w_{k+})$ is a diagonal matrix containing the row sums of $W$. The precision matrix $Q(W, \rho)$ is invertible if $\rho\in[0,1)$, and the corresponding univariate full conditional distribution for $\phi_{k}|\bd{\phi}_{-k}$ (where $\bd{\phi}_{-k}=(\phi_{1},\ldots,\phi_{k-1},\phi_{k+1},\ldots,\phi_{n})$) is given by

\begin{equation}
\phi_{k}| \bd{\phi}_{-k}~\sim~\mbox{N}\left(\frac{\rho\sum_{i=1}^{n}w_{ki}\phi_{i}}{\rho\sum_{i=1}^{n}w_{ki} + 1-\rho},~
\frac{\tau^{2}}{\rho\sum_{i=1}^{n}w_{ki} + 1-\rho}\right),\label{equation lerouxfc}
\end{equation}

where the conditioning is in fact only on the random effects in geographically adjacent areas (those having $w_{ki}=1$). An attractive feature of this model is that when $\rho=0$ the random effects are independent with a constant variance, while when $\rho=1$ the intrinsic model proposed by \cite{besag1991} is obtained. The partial correlations between $(\phi_{k},\phi_{j})$ implied by this model are given by

\begin{equation}
\corr{\phi_{k},\phi_{j}|\bd{\phi}_{-kj}}~=~\frac{\rho w_{kj}}{\sqrt{(\rho\sum_{i=1}^{n}w_{ji} + 1-\rho)(\rho\sum_{i=1}^{n}w_{ki} + 1-\rho)}}\label{equation partialcorrelation},
\end{equation}

which is zero for non-neighbouring areas. The equation also shows that if there is substantial spatial correlation between the majority of pairs of random effects (i.e. if $\rho$ is estimated as close to one), then all pairs of random effects that share a common border (have $w_{kj}=1$) will be correlated, and the strength of the partial correlation will depend on the number of other neighbouring areas. This is a reasonable assumption if the surface to be modelled has a single level of spatial smoothness, because the more neighbours a pair of areas have the less strongly they are related to each other conditional on their remaining neighbours. However, in the more realistic situation of having sub-regions of spatial smoothness separated by boundaries this model is not appropriate, because it will smooth over the random effects across the boundaries, resulting in poorer estimation of disease risk.\\

Inference for the parameters $\bd{\Theta}=(\bd{\phi}, \bd{\beta}, \rho, \tau^{2})$  is most commonly based on Markov Chain Monte Carlo (MCMC) simulation, although Integrated Nested Laplace Approximations (INLA) can also be used (for example  \cite{schrodle2011}). A number of software packages are available that fit the hierarchical model outlined above using MCMC simulation, including WinBUGS (\cite{lunn2000}), BayesX (\cite{belitz2009}) and the \emph{CARBayes} package in the statistical software  \texttt{R}, (\cite{R2009}) .

\subsection{Extensions to localised spatial smoothing}
The majority of research that has extended the  class of CAR priors to account for localised spatial smoothing has treated the neighbourhood matrix $W$ as a collection of additional random quantities, rather than being fixed and based on geographical adjacency. Specifically, the set of $\{w_{kj}|k\sim j\}$ are treated as binary random quantities, where as the set $\{w_{kj}|k\nsim j\}$ remain fixed at zero. Treating the former as random allows $(\phi_{k},\phi_{j})$ to be conditionally independent ($w_{kj}=0$) or correlated ($w_{kj}=1$), and if $w_{kj}$ is estimated as zero  a boundary is said to exist between the two random effects. One of the first models in this vein was developed by \cite{lu2007}, who proposed the logistic regression model 

\begin{eqnarray}
w_{kj}&\sim&\mbox{Bernoulli}(p_{kj})~~~~\forall~ k\sim j,\nonumber\\
\mbox{logit}(p_{kj})&=&\alpha_0 + \alpha_1z_{kj},\nonumber
\end{eqnarray}

where $z_{kj}$ is a non-negative quantity summarising the level of dissimilarity between areas $(A_{k}, A_{j})$. To overcome the weak identifiability of these parameters an informative prior is specified for  $\alpha_{1}$, although as argued by \cite{li2011}, such prior knowledge is unlikely to be available for a regression parameter that relates a  measure of dissimilarity to the spatial structure of the random effects. A similar approach was proposed by \cite{ma2010}, who replace the logistic regression model with either a second stage CAR prior or an Ising model. However, this approach requires tuning parameters to be specified by the user, the values of which appear to affect the results (see Tables 1 and 2 in the on-line appendix accompanying the paper). A related approach to that of \cite{lu2007} was proposed by \cite{lee2012}, who collectively model the set  $\{w_{kj}|k\sim j\}$ as a function of measures of dissimilarity, rather than modelling each element individually. However, while this approach largely overcomes the parameter identifiability problems encountered by \cite{lu2007}, it is restrictive in the sense that if $z_{tu}>z_{rs}$ then $\mathbb{P}(w_{tu}=1|\mathbf{Y}) <\mathbb{P}(w_{rs}=1|\mathbf{Y})$, no deviation from the ordering of the dissimilarity measures $\{z_{kj}\}$ is allowed.\\

An alternative approach was suggested by \cite{lee2012b}, who propose an iterative algorithm that re-estimates $W$ and the remaining model parameters  in turn, conditional on the current value of the other. Their algorithm is implemented using Integrated Nested Laplace Approximations (INLA, \cite{rue2009}) rather than MCMC for computational speed, and has the drawback that only an estimate of each $w_{kj}$ is provided, rather than the posterior probability that $w_{kj}=1$. A further alternative was proposed by \cite{li2011}, who  consider different $W$ matrices as different models, and use the Bayesian Information Criterion (BIC) to choose between them. However, the number of possible models is $2^{\mathbf{1}^TW\mathbf{1}/2}$, and to get around the large model space they only consider models that have one boundary, as part of a data mining technique.

%%%%%%%%%%
%%%% Methods
%%%%%%%%%%
\section{Methods}
We propose a two-stage approach to simultaneously estimating the model parameters $\bd{\Theta}$ and the spatial structure of the random effects $\{w_{kj}|k\sim j\}$, the first of which elicits a set of informative prior distributions, $\{\mathbb{P}(w_{kj}=1)|k\sim j\}$, while the second jointly estimates both sets of parameters using MCMC simulation. The motivation for an informative prior is that $\{w_{kj}|k\sim j\}$ are likely to be only  weakly identifiable from the single set of spatial data $\mathbf{Y}$, because they are large in number, and control the correlation between the random effects rather than being in the mean function for $\mathbf{Y}$. The likely weak identifiability of these parameters has also been discussed by \cite{li2011}, and is borne out empirically by the simulation study presented in Section 4 of this paper.

\subsection{Stage 1 - Eliciting informative priors for $\{w_{kj}|k\sim j\}$}
Eliciting prior information about the spatial structure of the random effects is not straightforward, because while one may have qualitative prior knowledge about the study region, for example about its underlying deprivation structure, using that to create $\{\mathbb{P}(w_{kj}=1)|k\sim j\}$ is far from straightforward. Therefore, we propose obtaining this prior information from a second data set, which comprises the same disease count data but for an earlier time period. Disease count data are typically available for multiple time periods, and unless large-scale urban re-generation has been undertaken, the spatial structure in these earlier data should be similar to that of the response. We note that in practice one could use the response variable for this purpose rather than earlier data, but this uses the data twice and is not in keeping with  the notion of prior beliefs.\\

Consider a vector of prior information $\bd{\phi}^{*}=(\phi_{1}^{*},\ldots,\phi_{n}^{*})$, which represents the spatial structure in the random effects surface for an earlier time period. For example, if no covariates are included in (\ref{equation likelihood}) then the random effects quantify the overall spatial structure in the log risk surface, and $\phi_{k}^{*}=\ln(Y_{k}^{*}/E_{k}^{*})$, the natural log of the ratio of the observed and expected disease count for area $k$ from an earlier time period. In contrast, in the presence of covariates the random effects represent the spatial pattern in the residual surface, in which case $\phi_{k}^{*}=\ln(Y_{k}^{*}/E_{k}^{*}) - \mathbf{x}_k^{\tiny\mbox{T}\normalsize}\hat{\bd{\beta}}$. The aim is to use these data to elicit $\{\mathbb{P}(w_{kj}=1)|k\sim j\}$, the set of probabilities that each pair of adjacent random effects are spatially correlated.  Two commonly used global measures of spatial correlation for the entire region are Geary's C (\cite{geary1954}) and Moran's I (\cite{moran1950}), which are given by

$$\mbox{Geary's C }= \frac{(n-1)\sum_{k\neq j}w_{kj}(\phi_{k}^{*}-\phi_{j}^{*})^{2}}{2[\sum_{k\neq j}w_{kj}]\sum_{i=1}^{n}(\phi_{i}^{*}-\bar{\phi^{*}})^{2}},$$

$$\mbox{Moran's I }= \frac{n\sum_{k\neq j}w_{kj}(\phi_{k}^{*}-\bar{\phi^{*}})(\phi_{j}^{*}-\bar{\phi^{*}})}{[\sum_{k\neq j}w_{kj}]\sum_{i=1}^{n}(\phi_{i}^{*}-\bar{\phi^{*}})^{2}}.$$

These statistics are weighted averages of the similarity of the response between all pairs of adjacent areas, and capture the level of spatial correlation globally across the entire region. However, as the aim is to quantify each $\mathbb{P}(w_{kj}=1)$ separately, natural statistics are the individual ordinates $(\phi_{k}^{*}-\phi_{j}^{*})^{2}$ and  $(\phi_{k}^{*}-\bar{\phi^{*}})(\phi_{j}^{*}-\bar{\phi^{*}})$. We elicit $\mathbb{P}(w_{kj}=1)$ by comparing the relative size of these ordinates for $(\phi_{k}^{*}, \phi_{j}^{*})$ to a reference distribution, which is the set of the $^{n}\mathcal{C}_{2}$ values created from all possible pairs of $(\phi_{r}^{*}, \phi_{s}^{*})$ over the study region. Thus, the reference distributions for the Geary's C and the Moran's I ordinates are

\begin{eqnarray}
\mathcal{P}_{G}&=&\{(\phi_{r}^{*}-\phi_{s}^{*})|1\leq r < s \leq n)\},\nonumber\\
\mathcal{P}_{M}&=&\{(\phi_{r}^{*}-\bar{\phi^{*}})(\phi_{s}^{*}-\bar{\phi^{*}})|1\leq r < s \leq n)\}.\nonumber
\end{eqnarray}

The set of prior probabilities are then computed as

\begin{eqnarray}
\mathbb{P}_{G}(w_{kj}=1)&=&\frac{|\{x\in\mathcal{P}_{G}|x>(\phi_{k}^{*}-\phi_{j}^{*})^{2}\}|}{|\mathcal{P}_{G}|},\nonumber\\
\mathbb{P}_{M}(w_{kj}=1)&=&\frac{|\{x\in\mathcal{P}_{M}|x<(\phi_{k}^{*}-\bar{\phi^{*}})(\phi_{j}^{*}-\bar{\phi^{*}})\}|}{|\mathcal{P}_{M}|},\nonumber
\end{eqnarray}

the proportion of the values from the reference distribution that correspond to bigger differences than that observed for $(\phi_{k}^{*}, \phi_{j}^{*})$. If random effects from neighbouring areas are spatially correlated they will have similar values, and  $\mathbb{P}_{M}(w_{kj}=1)$ and $\mathbb{P}_{G}(w_{kj}=1)$ should be close to one because the sets $(\mathcal{P}_{M}, \mathcal{P}_{G})$ contain large numbers of bigger values from geographically distant and hence uncorrelated areas. In contrast, if a neighbouring pair of random effects are very different, then $\mathbb{P}_{M}(w_{kj}=1)$ and $\mathbb{P}_{G}(w_{kj}=1)$ will be much closer to zero, suggesting a higher probability of there being a boundary between them.

\subsection{Stage 2 - overall hierarchical model}
Combining the elicited informative prior for $\{w_{kj}|k\sim j\}$ with the hierarchical model outlined by (\ref{equation likelihood}) and (\ref{equation lerouxfc}) gives the following joint model for $(\bd{\Theta},W)$:

\begin{eqnarray}
Y_{k}|E_k,R_k &\sim&\mbox{Poisson}(E_kR_k)~~~~\mbox{for }k=1,\ldots,n,\nonumber\\
\ln(R_{k})&=&\mathbf{x}_{k}^{\tiny\mbox{T}\normalsize}\bd{\beta} + \phi_{k},\nonumber\\
\beta_i&\sim&\mbox{N}(0, 1000),\nonumber\\
\phi_{k}| \bd{\phi}_{-k}&\sim&\mbox{N}\left(\frac{\rho\sum_{i=1}^{n}w_{ki}\phi_{i}}{\rho\sum_{i=1}^{n}w_{ki} + 1-\rho},~
\frac{\tau^{2}}{\rho\sum_{i=1}^{n}w_{ki} + 1-\rho}\right),\label{equation full model}\\
\tau^{2}&\sim&\mbox{Uniform}(0, 1000),\nonumber\\
\rho&\sim&\mbox{Uniform}(0, 1),\nonumber\\
w_{kj}&\sim&\mbox{Bernoulli}(p_{kj})~~~~\mbox{for }k\sim j.\nonumber
\end{eqnarray}

Here $p_{kj}$ is equal to either $\mathbb{P}_{G}(w_{kj}=1)$ or $\mathbb{P}_{M}(w_{kj}=1)$, depending on whether Geary's C or Moran's I ordinates are being used. Inference for this model uses MCMC simulation, and can be implemented in the  \emph{CARBayes} package in the statistical software  \texttt{R}. We note that this software can also fit models with Gaussian and binomial responses.

\subsection{Modelling uses}
The model described above can be used in two distinct ways, depending on the goal of the analysis. If the goal is to explain the spatial pattern in the response, then covariates should be included in the linear predictor, and the random effects will capture the residual spatial structure. Note, in this case the boundaries that are identified in the random effects surface do not correspond to boundaries in the disease risk surface. In contrast, if the goal is to identify boundaries in the risk surface, then the model can be fitted without covariates, because this ensures the random effects and risk surfaces have the same spatial structure, as $R_k=\exp(\beta_0+\phi_k)$. In this case $\rho$ should be fixed close to one, such as 0.99, so that globally the random effects are spatially smooth, with a boundary occurring between $(\phi_{k}, \phi_{j})$ if $w_{kj}=0$. We do not set $\rho=1$ in this scenario because the precision matrix $Q(W, \rho=1)$ would be singular and the univariate conditional distribution $\phi_{k}| \bd{\phi}_{-k}$ would be undefined if $\sum_{j=1}^{n}w_{kj}=0$. To assess the utility of using an informative prior for $\{w_{kj}|k\sim j\}$ we compare it to the following weakly informative alternatives, which allow the data to speak for themselves. 

\begin{enumerate}
\item[\textbf{Prior A}] - $f(\{w_{kj}|k\sim j\})=\prod_{k\sim j}\mbox{Bernoulli}(w_{kj}|0.5)$.  

\item[\textbf{Prior B}] - $f(\{w_{kj}|k\sim j\}, \alpha)=\prod_{k\sim j}\mbox{Bernoulli}(w_{kj}|\alpha)\times\mbox{Uniform}(\alpha|0,1)$. 

\item[\textbf{Prior C}] - $f(\{w_{kj}|k\sim j\}, \bd{\alpha}=\{\alpha_{kj}|k\sim j\})=\prod_{k\sim j}\mbox{Bernoulli}(w_{kj}|\alpha_{kj})$\\
$\times\prod_{k\sim j}\mbox{Uniform}(\alpha_{kj}|0,1)$. 
\end{enumerate}

Prior model A shows no preference for each $w_{kj}$ equalling one or zero, although across the complete set the expectation is that half would equal one. Prior model B relaxes this constraint, and instead allows the proportion of $\{w_{kj}|k\sim j\}$ that equal one to be estimated from the data. Prior model C relaxes this still further, and allows each $w_{kj}$ to have its own probability of equalling one. Preliminary analyses using simulated data showed that Priors A and C produced almost identical results, while Prior B performed badly by identifying almost no boundaries in all cases. The latter occurs because in the simulated data there were fewer boundaries than non-boundaries, which resulted in a low value of $\alpha$ thus making boundaries hard to identify. Therefore, the only weakly informative prior considered further in this paper is Prior A.

%%%%%%%%%%
%%%% Simulation
%%%%%%%%%%
\section{Simulation study}
This section presents a simulation study, which compares the performance of the global smoothing model proposed by \cite{leroux1999}, as well as the local smoothing model described in Section 3.2. The latter is applied with three different prior distributions for $\{w_{kj}|k\sim j\}$, which include those elicited from earlier data using Geary's C and Moran's I ordinates, as well as the weakly informative Prior A  model. This choice of models enable us to examine the effects of global versus local smoothing, as well as the effectiveness of using informative or weakly informative priors for the spatial structure of the random effects.

\subsection{Data generation and study design}
Simulated data are generated for the $n=271$ Intermediate Geographies (IG) that comprise the Greater Glasgow and Clyde health board, which is the study region used in the case study in Section 5. The disease counts are generated from model (\ref{equation likelihood}), where the expected numbers $\mathbf{E}$ are the emergency admissions data used in the case study.  The log risk surface is represented  by a single covariate and a set of random effects, the former being realisations of independent and identically distributed standard normal random variables. The corresponding regression parameter for this covariate is $\beta=0.1$, which is fixed for all simulated data sets and scenarios. In contrast, a new set of random effects are generated for each simulated data set, so that the results are not affected by a particular realisation of $\bd{\phi}$. Each scenario in this study is based on 200 simulated replicate data sets, each of which consists of  two sets of random effects $(\bd{\phi}, \bd{\phi}^{*})$, which are used to create corresponding response vectors  $(\mathbf{Y}, \mathbf{Y}^{*})$. The first of these is the response variable while the second is the data from an earlier time period, and they have a correlation of 0.95 which corresponds to the value observed for the emergency admissions data analysed in the next section. We note that our proposed method would not be appropriate if the correlation between the earlier data and the response was weak.\\

\begin{figure}
\centering\caption{Locations of the boundaries (bold black lines) in the simulated random effects surfaces.}
\label{figure simstudy}\scalebox{0.4}{\includegraphics{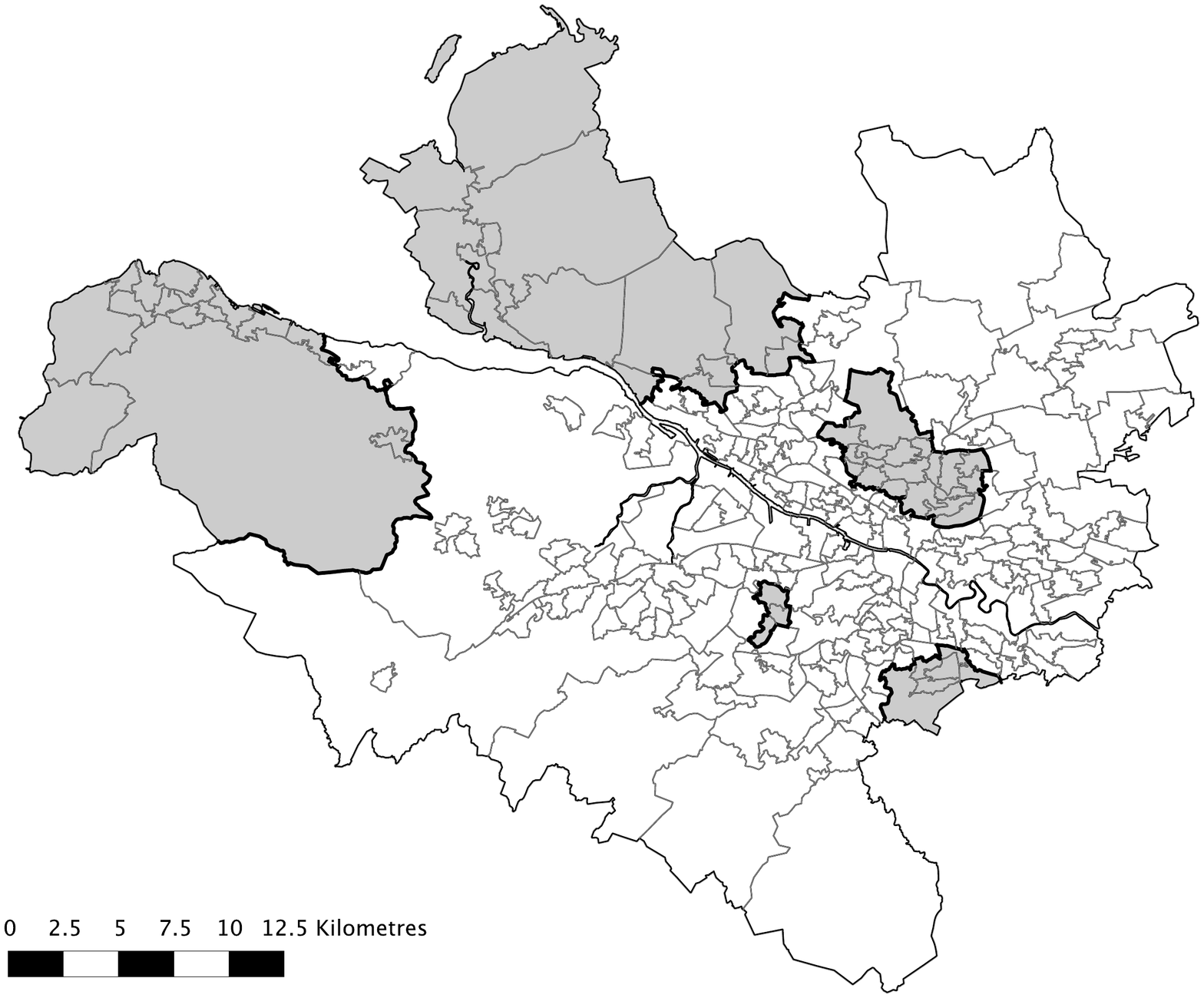}}
\end{figure}

Localised spatial structure is induced into the random effects by generating them from a multivariate Gaussian distribution, where the mean is piecewise constant, while the covariance matrix is specified by the spatially smooth Matern correlation function. For the latter, the spatial smoothness parameter is fixed at 2.5 (so that the surface is twice mean square differentiable), while the spatial range is chosen so that the median correlation between all pairs of areas is 0.5. The localised spatial smoothness follows the template shown in Figure \ref{figure simstudy}, where the expectation of $\bd{\phi}$ is equal to zero for the  large white region and equal to $M$ in the 5 smaller grey regions. The latter represent the locations of high risk clusters, and thus boundaries in the random effects surface occur where white and grey regions border each other (the bold lines in Figure \ref{figure simstudy}). This template results in 74 boundaries, which is around 10$\%$ of the total number of boundaries in the study region (701). The values of $M$ used in this study are $M=0.5$ and $M=1$, with the larger value corresponding to greater differences between the random effects in these areas (larger boundaries).

\subsection{Results - Prior elicitation}
The performance of the prior elicitation using the Geary's C and the Moran's I ordinates is summarised in Table  \ref{table boundary}. The table displays both the sensitivity, the proportion of times a boundary is correctly identified, and the specificity, the proportion of times a non-boundary is correctly identified. The sensitivity is calculated by computing the proportion of the true boundaries that have $\mathbb{P}(w_{kj}=0)>c$, where $c=0.5, 0.75, 0.9$, while the specificity is the proportion of the non-boundaries that have $\mathbb{P}(w_{kj}=1)>c$, for the same values of $c$. For each simulated data set the prior elicitation approach is applied to the residuals $\hat{\phi}_{k}^{*}=\ln(Y_{k}^{*}/E_{k})-x_{k}\hat{\beta}$, which result from applying a Poisson generalised linear model to the data from an earlier time period. Table \ref{table boundary} shows that both methods perform well relative to the weakly informative specification that $\mathbb{P}(w_{kj}=0)=0.5~\forall~k\sim j$, with the sensitivity and specificity at the lowest probability threshold of 0.5 ranging between 0.6 and 1.0 for both values of $M$. These  values reduce as the threshold level $c$ becomes more stringent as expected, but they remain between 0.39 and 0.77 at the 0.75/0.25 threshold. Finally, the results from the Geary's C ordinates almost always outperform the Moran's I values, with differences as large as 0.2. In contrast, in the few cases where Moran's I is superior the differences are less than 0.05.

\subsection{Results - Posterior inference}
The performance of the four models is summarised in Tables \ref{table boundary} and \ref{table bias}, which summarise the boundary identification and estimation performance respectively.  Table \ref{table bias} displays the bias and root mean square error (RMSE) for $\beta$ and $\mathbf{R}=(R_{1},\ldots,R_{n})$, where both quantities are presented as percentages of their true values. All models produce close to unbiased results for $(\beta, \mathbf{R})$ for both values of $M$, with percentage biases being less than 0.5$\%$ in all cases. However, in the presence of localised spatial structure the global Leroux model performs worse in terms of RMSE than the three local models for both $\beta$ and $\mathbf{R}$, with differences ranging between 0.66$\%$ and 3.53$\%$. The differences between the three local models are small for both values of $M$, and we do not over interpret these results.\\

In terms of boundary identification, Table \ref{table boundary} shows that the model using the weakly informative Prior A can identify a sizeable proportion of the boundaries, with sensitivities of around 0.86 (c=0.5), 0.55 (c=0.75) and 0.44 (c=0.9) respectively for both values of $M$. However, its specificity is very poor, with values of around  0.65 (c=0.5), 0.03 (c=0.25) and 0.01 (c=0.1). In contrast, the models with informative priors perform much better, with improvements of between 0.03 and 0.33  in sensitivity and 0.11 and 0.37 in specificity depending on $M$ and $c$. In common with the prior elicitation, the Geary's C model outperforms the Moran's I one in almost all cases, and these models do exhibit some prior to posterior learning.

\begin{table}
\caption{Sensitivity and specificity for the identification of boundaries based on the proportion of boundaries satisfying $\mathbb{P}(w_{kj}=0)>c$ and the proportion of non-boundaries  satisfying $\mathbb{P}(w_{kj}=1)>c$.} \label{table boundary}
\centering\begin{tabular}{llrrrrr}
\hline
 &\raisebox{-1.5ex}[0pt]{\textbf{Threshold}} &\multicolumn{2}{c}{\textbf{Prior elicitation}}&\multicolumn{3}{c}{\textbf{Posterior inference}}\\
&&\textbf{Geary}&\textbf{Moran} &\textbf{Geary}& \textbf{Moran} & \textbf{A}\\\hline

Sensitivity&c=0.5&1.00&0.80&0.98&0.96&0.86\\
M=1&c=0.75&0.77&0.66&0.82&0.81&0.55\\
&c=0.9&0.23&0.25&0.77&0.75&0.44\\\hline

Sensitivity&c=0.5&0.79&0.59&0.93&0.89&0.86\\
M=0.5&c=0.75&0.44&0.38&0.72&0.64&0.54\\
&c=0.9&0.14&0.13&0.64&0.55&0.43\\\hline

Specificity&c=0.5&0.89&0.77&0.84&0.80&0.66\\
M=1&c=0.75&0.51&0.46&0.40&0.37&0.03\\
&c=0.9&0.21&0.26&0.21&0.23&0.01\\\hline

Specificity&c=0.5&0.80&0.73&0.78&0.76&0.65\\
M=0.5&c=0.75&0.45&0.42&0.36&0.35&0.02\\
&c=0.9&0.19&0.23&0.18&0.21&0.01\\\hline
\end{tabular}
\end{table}

\begin{table}
\caption{Bias and root mean square error (RMSE) of $\beta$ and the fitted risk surface $\mathbf{R}$ for each of the four models. All results are presented as percentages of their true values.} \label{table bias}
\centering\begin{tabular}{llrrrr}
\hline
\raisebox{-1.5ex}[0pt]{\textbf{Metric}}  &\raisebox{-1.5ex}[0pt]{\textbf{M}} &\multicolumn{4}{c}{\textbf{Model}}\\
&&\textbf{Leroux}&\textbf{Geary}& \textbf{Moran} & \textbf{A}\\\hline

Bias $\beta$&1&0.309&0.195&0.163&0.425\\
&0.5&0.245&0.344&0.087&0.206\\\hline
RMSE $\beta$&1&4.862&1.464&1.558&1.296\\
&0.5&2.682&1.405&1.531&1.287\\\hline

Bias $\mathbf{R}$&1&-0.058&-0.026&-0.027&-0.025\\
&0.5&-0.095&-0.048&-0.051&-0.043\\\hline
RMSE $\mathbf{R}$&1&4.074&3.253&3.372&3.217\\
&0.5&4.289&3.431&3.631&3.387\\\hline
\end{tabular}
\end{table}

%%%%%%%%%%
%%%% Application
%%%%%%%%%%
\section{Case study}
We illustrate the methodology by mapping the spatial pattern in the risk of emergency admission to hospital in Greater Glasgow, Scotland in 2010, which will be of interest to health professionals who deal with resource allocation. We address two different epidemiological questions in this study: (i) what is the spatial pattern in respiratory disease risk and what factors affect it? and (ii) where and how many boundaries are there in the estimated risk surface? 

\subsection{Data description}
\begin{figure}
\centering\caption{Standardised Incidence Ratio (SIR) for emergency admission to hospital.}
\label{figure SIR}\scalebox{0.4}{\includegraphics{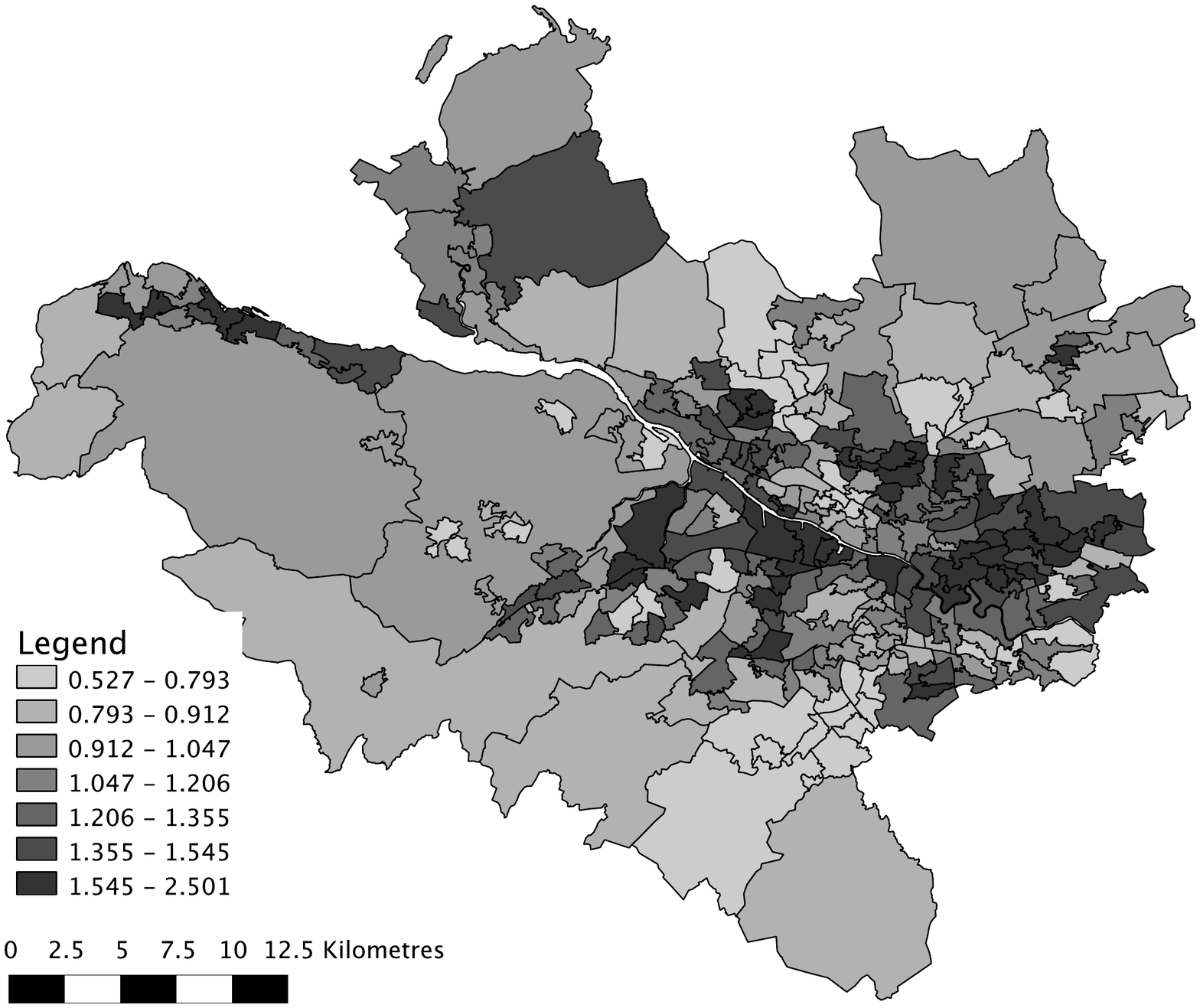}}
\end{figure}

The study region is the health board comprising the city of Glasgow and the river Clyde estuary,  which in 2010 contained just under 1.2 million people. For the purposes of this study it is partitioned into $n=271$ Intermediate Geographies (IG), which are administrative units constructed based on population size rather than geographical area (on average just over 4,000 people live in each area). The data used in this study come from the Scottish Neighbourhood Statistics (SNS) database (\emph{http://www.sns.gov.uk}), and the response variable is the number of emergency admissions to non-psychiatric and non-obstetric hospitals in each IG in 2010. The data used to elicit the set of priors for $\{w_{kj}| k\sim j\}$ is the same variable but for 2009, and in both cases the expected numbers of admissions were calculated by external standardisation, using age and sex specific rates for the whole of Scotland. The SIR is displayed for the 2010 data in  Figure \ref{figure SIR}, which shows that the risks are highest in the heavily deprived east end of Glasgow (east of the study region), as well as along the southern bank of the river Clyde, both in and out-with the city. The correlation between the SIR values for the two years is 0.95, suggesting that the 2009 data should be appropriate for eliciting prior probabilities for the spatial structure of the 2010 data.\\

We consider three potential covariates in this study, which are likely to affect the risk of emergency admission to hospital. The first of these is a measure of socio-economic deprivation (\cite{reid1999}), specifically, the percentage of people in each IG that are defined to be income deprived (are in receipt of a combination of means-tested benefits). The second is a measure of ethnicity, because it is likely that the level of risk inducing behaviour (such as excessive consumption of alcohol) will differ between ethnic groups. The only variable available to measure ethnicity is the percentage of school children in each IG from ethnic minorities (non-white), which we appreciate is imperfect in many ways (for example, it does not differentiate between different ethnic groups).  Finally, a variable measuring the time taken to drive to a doctors surgery is also available, which may affect their decision as to whether to go to hospital.

\subsection{Modelling}
Initially, a model including all three covariates but without the random effects was fitted to the data, to determine the existence of any residual spatial correlation. However, the ethnicity covariate is highly skewed to the right, as the majority of the intermediate geographies have a very small (or zero) percentage of people who are non-white. Therefore, a log transform of this covariate was used instead, where a constant of 0.5  was added to prevent the occurrence of $\ln(0)$. The adequacy of this covariate only model was then assessed, and substantial overdispersion was found (overdispersion parameter equal to 8.27), as well as spatial correlation in the residuals. The latter was assessed by computing Geary's C (C = 0.7769) and Moran's I (I = 0.2398) statistics, and both showed substantial evidence of residual spatial correlation. To alleviate these problems random effects were added to the model, and we compare the four models used in the simulation study, where the informative priors are elicited from the 2009 data.\\

In addition, the random effects models listed above are also applied to the data without the covariates, because this enables the number and locations of any boundaries in the risk surface to be identified. Recall, that if covariates are included in the model then the boundaries only relate to the random effects surface, and do not relate to the risk surface. Posterior inference for all models is based on 3 parallel Markov chains, which were burnt in for 50,000 iterations (by which time convergence was assessed to have been reached) and then run for an additional 50,000 iterations, yielding 150,000 samples in total. The goodness-of-fit of these models is summarised by the Deviance Information Criterion (DIC, \cite{spiegelhalter2002}), where smaller values represent a better fitting model. Without the covariates the DIC values were: Leroux - 2690.5, Prior A - 2689.6, Prior Geary - 2674.0, and Prior Moran - 2680.7, suggesting that the global Leroux model provides the worst fit to the data, although the difference between it and the local model with a weakly informative prior is very small. In contrast, the local spatial smoothing models with priors elicited using the Geary's C and the Moran's I ordinates fit the data much better, with DIC values that are smaller than the Leroux prior by 15.5 and 9.8 respectively. The inclusion of the covariates reduced the DIC values of all four models by around 25, but their relative sizes remain unchanged.

\subsection{Results}

\subsubsection{Covariate effects}
The covariate effects are shown in Table \ref{table results2}, which displays posterior medians as well as 95$\%$ credible intervals. All results are presented on the relative risk scale, for a one standard deviation increase in each covariates value. There is strong evidence that income deprivation affects emergency admission to hospital, with populations exhibiting nearly 13$\%$ higher levels of income deprivation having between  a 27$\%$ and a 29$\%$ increased risk. The proportion of the population who are non-white also affects emergency admissions, with higher proportions having smaller risks by around 3$\%$. In contrast, the time taken to drive to a GP's surgery does not appear to affect emergency admission to hospital, as the 95$\%$ credible intervals for all models include the null risk of one. The estimated relative risks for the income deprivation and drive time covariates differ between the global Leroux and the local models, and the simulation study suggests that the latter estimates are likely to be more accurate.

\begin{table}
\caption{\label{table results2} Estimated covariate effects (posterior medians) and 95$\%$ credible intervals for a one standard deviation increase (listed in row 1) in each covariate.}
\centering
\small
\begin{tabular}{lrrr}
\hline \textbf{Model}&\multicolumn{3}{c}{\textbf{Covariate effects}}\\
&\textbf{Drive time}&\textbf{Ethnicity}& \textbf{Income deprivation}\\\hline
\textit{SD}&\textit{0.715}&\textit{1.005}&\textit{12.75}\\

Leroux&0.987 (0.966, 1.007)&0.969 (0.949, 0.991)&1.277 (1.255, 1.302)\\
Prior  Geary&0.991 (0.974, 1.008)&0.968 (0.948, 0.987)&1.282 (1.260, 1.304)\\
Prior Moran&0.991 (0.972, 1.011)&0.970 (0.950, 0.991)&1.282 (1.259, 1.304)\\
Prior A&0.991 (0.971, 1.011)&0.969 (0.947, 0.989)&1.280 (1.255, 1.302)\\
\hline
\end{tabular}
\normalsize
\end{table}

\subsubsection{Estimated risk maps}
\begin{figure}
\centering\caption{Maps displaying the estimated spatial pattern in disease risk and the location of the boundaries. The top panel displays the estimate from the elicited prior using Geary's C ordinates, while the bottom one relates to the weakly informative prior model A. The white lines denote the locations of the boundaries, as defined by having $\mathbb{P}(w_{kj}=0|\mathbf{Y})>c$, with $c=0.5$ (thin line), $c=0.75$ (medium line) and $c=0.9$ (thick line).}\label{figure results}
\begin{picture}(20,20.5)
\put(0,2){\scalebox{0.45}{\includegraphics{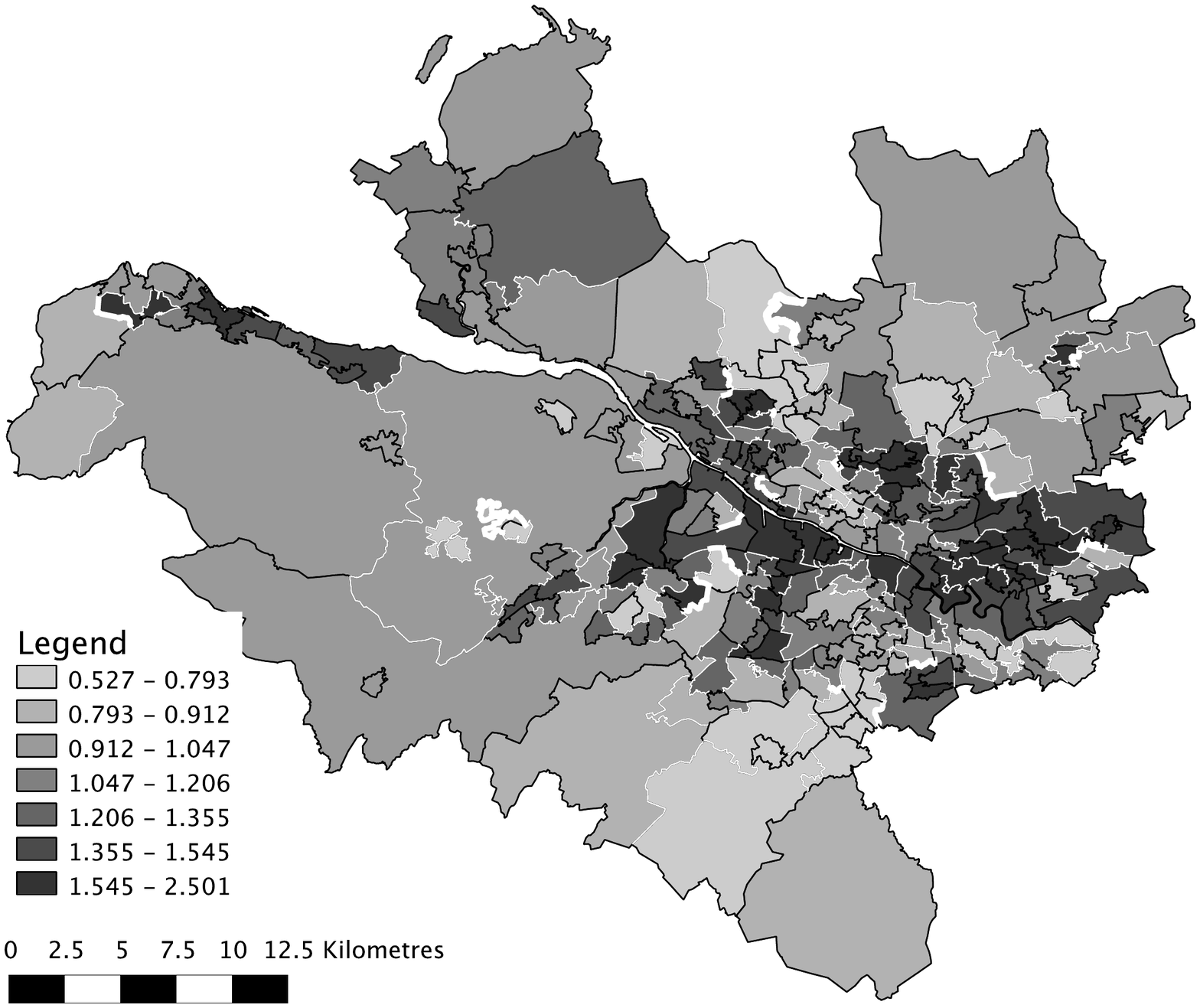}}}
\put(0,11){\scalebox{0.45}{\includegraphics{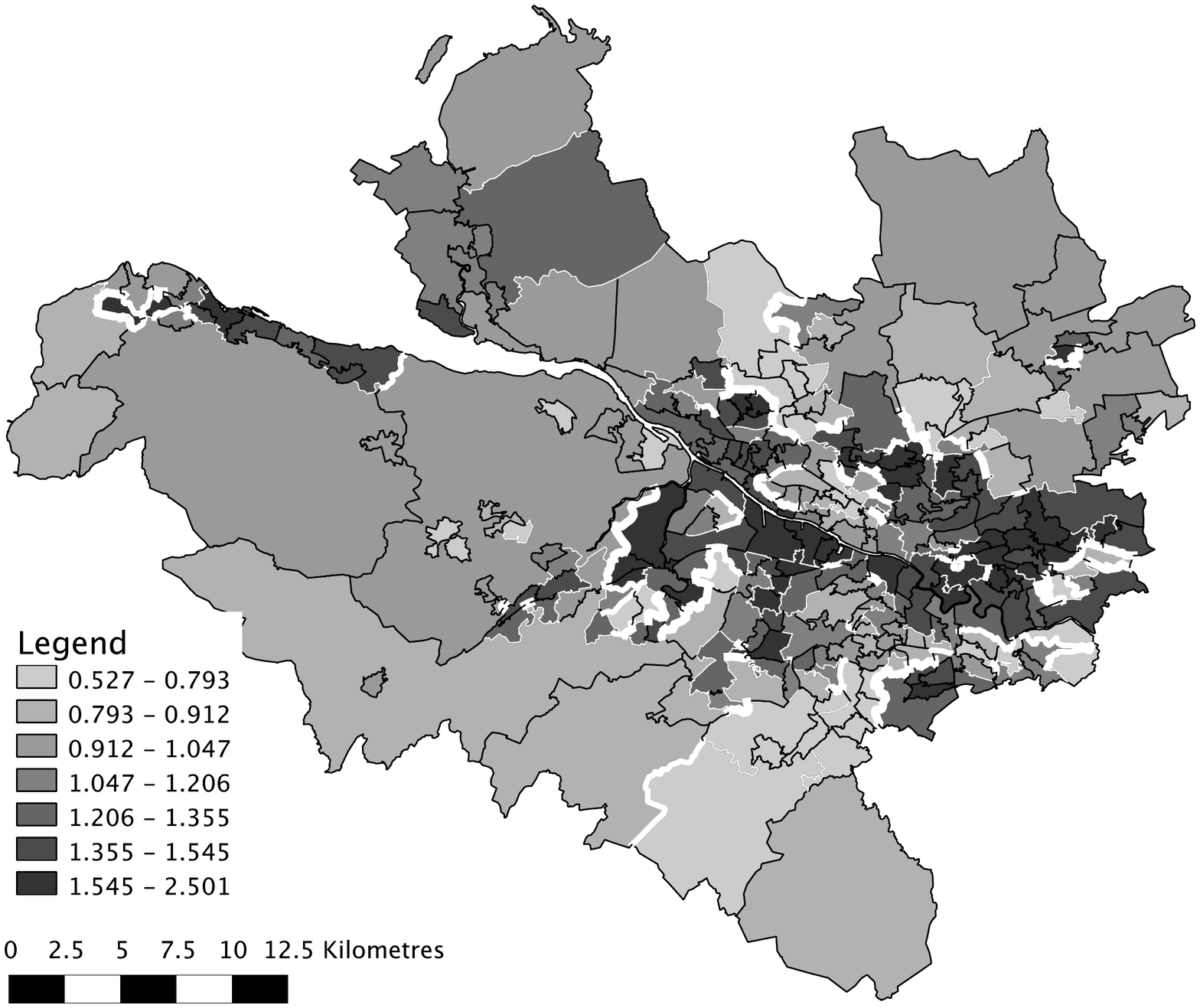}}}
\end{picture}
\end{figure}

The estimated risk maps from two of the localised spatial smoothing models are presented in Figure \ref{figure results}, where the top panel displays the estimate from the Prior Geary model, while the bottom panel relates to the Prior A model. The results from the Prior Moran model are somewhat similar to those from the Prior Geary model, and for brevity are not shown. The scale on the maps is the same as that used in Figure \ref{figure SIR}, with higher risk areas having darker shading. Both maps display the results from models without covariate effects, but the estimated spatial risk patterns remain similar when covariates are included in the model. Qualitatively, the maps display the same spatial structure as the raw SIR values presented in Figure \ref{figure SIR} as expected, although they are spatially smoother and exhibit less extreme values. One of the most notable features of these maps is that they display largely the same spatial pattern as that of the income deprivation covariate (map not shown), with the affluent west end of Glasgow (middle of the map) being at very low risk, where as the deprived east and north of the city are at much higher risk. The mean risk across the Greater Glasgow region is 1.168, which suggests that on average Greater Glasgow has a 16.8$\%$ increased risk of emergency admission to hospital compared with the overall Scottish average.

\subsubsection{Boundary identification}
\begin{figure}
\centering\caption{Prior and posterior probabilities that each $w_{kj}=0$ from the Prior Geary, Prior Moran and Prior A  models.}
\label{figure boundaries}\scalebox{0.4}{\includegraphics{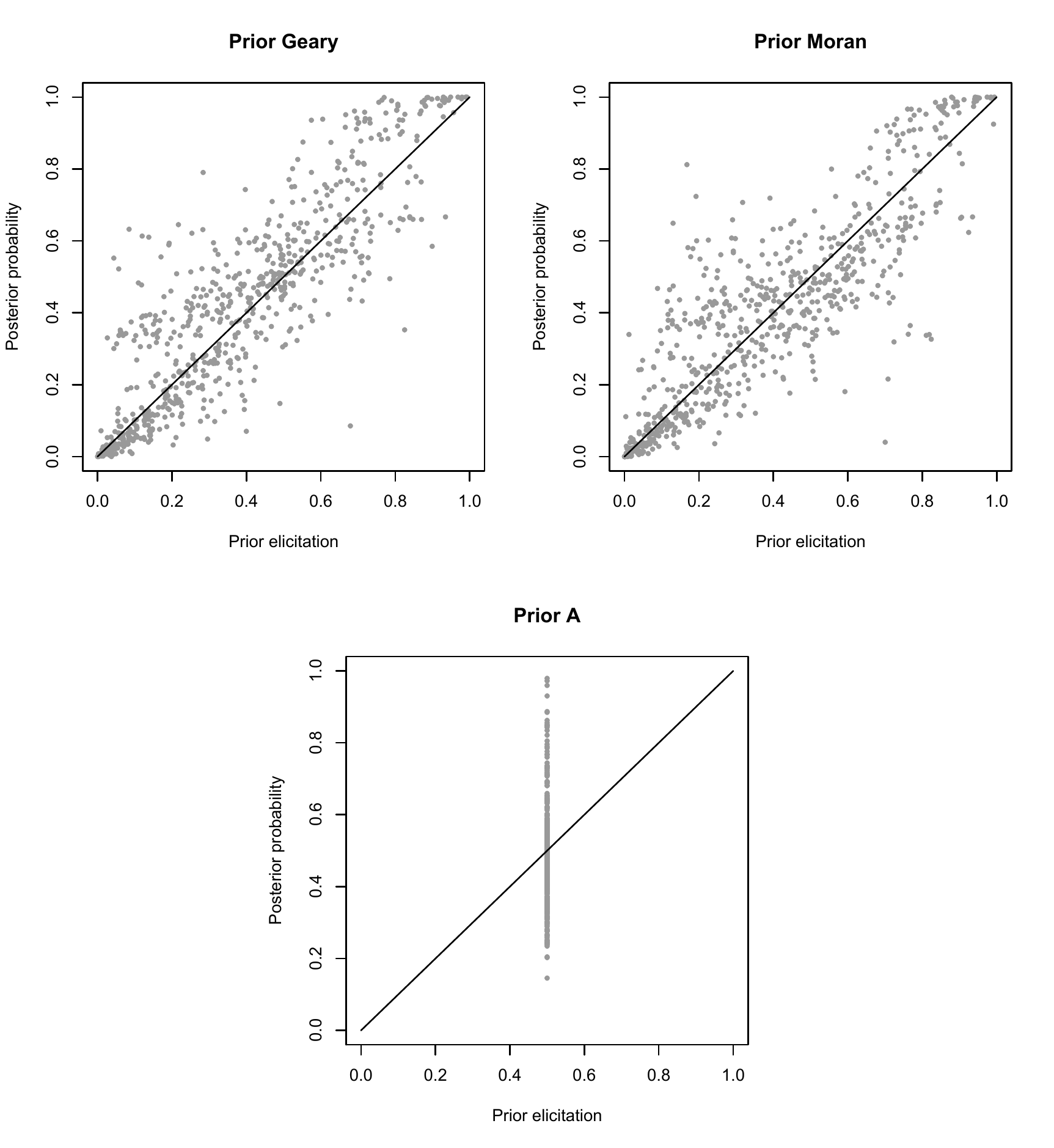}}
\end{figure}

Table \ref{table results1} and Figures \ref{figure results} and \ref{figure boundaries} summarise the numbers and locations of the boundaries that have been identified, both in the prior elicitation process and as a result of fitting model (\ref{equation full model}) with different priors. For brevity, only the results from the models without covariate effects are included, and the residual boundaries  corresponding  to models including covariates are not discussed. The table shows that the prior elicitation based on Geary's C and Moran's I ordinates has identified around 33$\%$ of the borders as boundaries based on the balance of probability (i.e c=0.5), while for the more stringent threshold of $c=0.75$ this reduces to around 10$\%$. At the $c=0.75$ threshold Geary's C has identified 69 boundaries compared to 81 for Moran's I, of which 61 of these have been identified by both ordinates. The differences between the prior elicitation and the posterior distributions for $w_{kj}$ for all 3 models  are displayed in Figure \ref{figure boundaries}, where in all cases the results presented are the prior and  posterior probability that each $w_{kj}$ equals zero. The line of equality is added to each plot to aid the presentation. The figure shows that there is some prior to posterior learning, because the points are not all concentrated on the line of equality. However, as would be expected the prior does have a sizeable impact on the posterior, as only 4 out of the 701 $\mathbb{P}(w_{kj}=0)$ values changed by more than 0.5 for both the Prior Geary and Prior Moran models.\\

The locations of the boundaries are displayed as white lines in Figure \ref{figure results}. The thickness of the line corresponds to the size of the posterior probability that a border is a boundary, i.e. to $\mathbb{P}(w_{kj}=0|\mathbf{Y})$. The thin lines correspond to posterior probabilities between 0.5 and 0.75, the medium lines to probabilities between 0.75 and 0.9, while the thick lines are probabilities greater than 0.9. The boundaries that have been identified by the Prior Geary and Prior A models are very different, both in their numbers and locations. Table \ref{table results1} shows that the weakly informative Prior A model has identified 41$\%$ of the borders as being a boundary if the threshold is the balance of probability (i.e. greater than 0.5), which decreases rapidly to only 3.1$\%$ when a 0.75 cutoff is used. In fact, for this model 95.3$\%$ of the posterior probabilities for $\{w_{kj}|k\sim j\}$ are between 0.25 and 0.75, suggesting that in the absence of prior information the data have largely been unable to distinguish between boundaries and non-boundaries. This is also illustrated by Figure \ref{figure results}, which shows large numbers of thin white lines (based on a 0.5 threshold) that do not appear to correspond to sizeable changes in the estimated risk surface.\\

In contrast, the results from the Prior Geary model appear to be much more reasonable, as the boundaries that have been identified correspond to sizeable changes in disease risk. This model has identified 37.2$\%$ of borders as boundaries at the 0.5 level, which drops to 13.7$\%$ at the 0.75 level. The addition of prior information has also managed to better distinguish between boundaries and non-boundaries, with only 53.9$\%$ of the posterior probabilities for $\{w_{kj}|k\sim j\}$ being between 0.25 and 0.75. The other interesting aspect of these results is that the majority of the boundaries are open, and thus do not completely enclose a region or set of regions. The identification of closed boundaries would be more interesting from a cluster detection viewpoint, because it enables sub-regions to be identified as having `different' values from all of their neighbours. However, Figure \ref{figure results} shows that the spatial structure in these disease data is far more complex, and forcing the boundaries to be closed would not provide an appropriate representation of this structure if the goal of the analysis is to quantify this structure. Finally, we note that areas on opposite banks of the river Clyde (the thin white line running south east) are not assumed to be neighbours, which explains the absence of boundaries in this area.

\begin{table}
\caption{\label{table results1} Summary of the numbers of boundaries identified in the risk surface (without covariates) from both prior elicitation and posterior inference}
\centering
\small
\begin{tabular}{llccc}
\hline 
&&\multicolumn{3}{c}{\textbf{Percentage of boundaries identified}}\\
\textbf{}&\textbf{Model}&\multicolumn{3}{c}{\textbf{based on $\mathbb{P}(w_{kj}=0|\mathbf{Y})>c$}}\\
&&$c=0.5$&$c=0.75$&$c=0.9$\\\hline
Prior&Prior Geary&32.1$\%$&9.8$\%$&2.7$\%$\\
elicitation&Prior Moran&34.3$\%$&11.6$\%$&2.7$\%$\\
&Prior A&0$\%$&0$\%$&0$\%$\\\hline

Posterior&Prior Geary&37.2$\%$&13.7$\%$&8.0$\%$\\
inference&Prior Moran&31.0$\%$&10.6$\%$&6.8$\%$\\
&Prior A&41.1$\%$&3.1$\%$&0.6$\%$\\\hline
\end{tabular}
\normalsize
\end{table}

%%%%%%%%%%
%%%% Application
%%%%%%%%%%
\section{Discussion}
This paper has highlighted the limitations of using global spatial smoothing models in a disease mapping context, which are overly restrictive because real disease data are likely to contain areas of smooth evolution as well as locations where sharp changes (boundaries) occur. This problem is overcome by  modelling the elements of the neighbourhood matrix $\{w_{kj}|k\sim j\}$ as binary random quantities, and this extension is relatively new in a disease mapping context, with Section 2.3 providing a brief review of the work to date in this area. The inherent problem in modelling the elements of the neighbourhood matrix is the large numbers of additional correlation parameters that need to  be estimated, which outnumber the data and are likely to be only weakly identifiable. The approach proposed here overcomes this problem by eliciting an informative prior for each $w_{kj}$, based on the well known Geary's C and Moran's I measures of spatial autocorrelation. The prior information is obtained from data relating to an earlier time period than the response, which should be straightforward to obtain due to the wide availability of small-area data. This approach has advantages over existing alternatives, such as the ease of eliciting prior information (compare with \cite{lu2007}), not requiring tuning parameters (see \cite{ma2010}), and being fully Bayesian (see \cite{lee2012b}).\\

The simulation study presented in Section 4 illustrates 3 main points. Firstly, in the presence of localised spatial structure, global smoothing models produce poorer estimates of covariate effects and disease risk (in terms of RMSE) compared to models that can capture localised spatial structure (see Table \ref{table bias}). Secondly, while using a weakly informative prior for $\{w_{kj}|k\sim j\}$ does not adversely affect the estimation of covariate effects and disease risk, it cannot accurately identify boundary locations. In particular, the majority of the posterior probabilities $\{\mathbb{P}(w_{kj}=0|\mathbf{Y})|k\sim j\}$ are close to 0.5, the prior value, and the sensitivity and especially the specificity at the more stringent thresholds are poor. Thirdly, the study shows that eliciting prior information improves the sensitivity and specificity at all thresholds, with the prior based on Geary's C outperforming the Moran's I prior in almost all cases. The case study presented in Section 5 largely reinforces these conclusions, as the Geary's C prior model provides the best fit to the data (measured via DIC) compared to the remaining models. In addition, the boundaries identified by this model appear to correspond to sizeable changes in the risk surface, a feature which is not the case when using a weakly informative prior for $\{w_{kj}|k\sim j\}$.\\

This paper suggests two natural avenues for future work. The first is the extension of these methods to  a spatio-temporal or multiple disease context, which could address the question of whether the locations of risk boundaries changed over time or between multiple diseases. A hidden Markov model for each $w_{kj}$ might be an appropriate in this context, which would control the evolution of $W$ over time or across diseases. The second extension is to force the boundaries to be closed, for example see \cite{knorrheld2000}, which would allow clusters of high risk areas to be singled out from their neighbours. This problem could be addressed by modelling $\{w_{kj}|k\sim j\}$ in the manner proposed here, but additional constraints would have to be enforced on this set to ensure the boundaries were always closed.

\section*{Acknowledgements}
This research was funded by the Economic and Social Research Council (RES-000-22-4256).

\bibliographystyle{chicago}
\bibliography{Lee}

\end{document}